\documentclass[conference]{IEEEtran}
\ifCLASSINFOpdf
 \usepackage[pdftex]{graphicx}
\else
\fi

\usepackage{cite}
\usepackage{amsmath,amssymb,amsfonts}
\usepackage{algorithmic}
\usepackage{graphicx}
\usepackage{textcomp}
\usepackage{subfigure}
\usepackage{mathrsfs}
\usepackage{threeparttable}
\usepackage{float}
\usepackage{epsfig}
\usepackage{epstopdf}
\usepackage{amsmath,bm}
\usepackage{amsmath}
\usepackage{multirow}
\usepackage{color, soul}
\soulregister\cite7 
\soulregister\citep7 
\soulregister\citet7 
\soulregister\ref7 
\soulregister\pageref7 

\def\BibTeX{{\rm B\kern-.05em{\sc i\kern-.025em b}\kern-.08em
    T\kern-.1667em\lower.7ex\hbox{E}\kern-.125emX}}
\usepackage{makecell}
\usepackage[bookmarksnumbered=true, bookmarksopen=true]{hyperref}

\begin{document}

\title{Path Loss Analysis for Low-Altitude Air-to-Air Millimeter-Wave Channel in Built-Up Area}
\author{\IEEEauthorblockN{Zhuangzhuang~Cui\IEEEauthorrefmark{2}, Abdul Saboor\IEEEauthorrefmark{2}, Achiel Colpaert\IEEEauthorrefmark{2}, and Sofie Pollin\IEEEauthorrefmark{2}}
\IEEEauthorblockA{\IEEEauthorrefmark{2}WaveCoRE, Department of Electrical Engineering (ESAT), Katholieke Universiteit Leuven, Belgium\\
Email: \texttt{\{zhuangzhuang.cui\}@kuleuven.be}}}



\maketitle

\begin{abstract}
Communications between unmanned aerial vehicles (UAVs) play an important role in deploying aerial networks. Although some studies reveal that drone-based air-to-air (A2A) channels are relatively clear and thus can be modeled as free-space propagation, such an assumption may not be applicable to drones flying in low altitudes of built-up environments. In practice, low-altitude A2A channel modeling becomes more challenging in urban scenarios since buildings can obstruct the line-of-sight (LOS) path, and multipaths from buildings lead to additional losses. Therefore, we herein focus on modeling low-altitude A2A channels considering a generic urban deployment, where we introduce the evidence of the small-size first Fresnel zone at the millimeter-wave (mmWave) band to approximately derive the LOS probability. Then, the path loss under different propagation conditions is investigated to obtain an integrated path loss model. In addition, we incorporate the impact of imperfect beam alignment on the path loss, where the relation between path loss fluctuation and beam misalignment level is modeled as an exponential form. Finally, comparisons with the 3GPP model show the effectiveness of the proposed analytical model. Numerical simulations in different environments and heights provide practical deployment guidance for aerial networks.
\end{abstract}


\section{Introduction}
Unmanned aerial vehicles (UAVs) are envisioned as an integral component of next-generation communication systems due to their low cost, high mobility, and rapid on-demand deployment. Prior studies have developed numerous use cases incorporating drones into wireless communications. For example, drones can be used as temporary aerial base stations serving ground users or vehicles in case the terrestrial infrastructure is unavailable \cite{x1}. In such scenarios, features such as rapid deployment, controllable height, and interconnectivity between multiple UAVs are essential. Therefore, a robust link between drones is needed to guarantee reliable communication in the aerial network, as a prerequisite 
for the design and optimization of the communication upper layers \cite{x2}. For civil applications, the emerging aerial network is promising to realize the prospect of a smart city, where the UAV can be used in the delivery, transportation, traffic, base station, relay, etc \cite{x3}. 
It shows that drone-to-drone communications are as important as air-to-ground (A2G) communications that have been under intensive research \cite{y1}. To facilitate aerial system research, the design of robust UAV commutation systems needs a better understanding of the corresponding air-to-air (A2A) channels, and accurate, general, and easy-to-use channel models. 

A selection of prior works discusses A2A channel modeling, and also poses some challenges for future research \cite{x4}. In \cite{x5}, A2A channel measurements were conducted with IEEE 802.11n (2.4~GHz) in altitudes below 50~m, where a height-dependent 
Ricean channel model was proposed. Measurements in an open field at the same frequency band were carried out in \cite{x6}, where the results showed that the path loss exponent (PLE) is 
approximating the free-space propagation as the flight height of UAV increases. Wideband measurements were conducted at 5.2 GHz in an urban environment \cite{x7}, where the buildings' reflections play an important role in the power delay profile (PDP). Moreover, for flights higher than buildings, the reflection from the rooftop is clearly observed in the PDP. Other related works also include multi-environmental A2A channel measurements at S-band and C-band conducted in \cite{x8}.

By employing a geometry-based stochastic model (GBSM), authors in \cite{x9} proposed a 3D ellipsoid channel model for A2A links, where scatterers were divided into local and clustering 
type, and power delay profiles were obtained by the superposition of two kinds of rays. Exploiting the prolate spheroidal coordinate system \cite{x10}, non-stationary A2A channels were analyzed, in terms of time-variant and delay-dependent Doppler frequency. Authors in \cite{x11} utilized a 3D cylinder GBSM and incorporated a Markov mobility model for UAV for low-altitude A2A channels. These models are superior in terms of scalability, however, need more validations with actual measurements. This paper shows a comprehensive structure of Fresnel zone-assisted channel modeling for A2A links. The main contributions are summarised as follows. \emph{First}, we propose a complete architecture for A2A channel large-scale fading analysis, including line-of-sight (LOS) probability modeling, and path loss modeling in both LOS and non-line-of-sight (NLOS) conditions. \emph{Second}, we introduce a universal urban model to describe the built-up area mathematically, which significantly improves the extensibility of proposed channel models. \emph{Third}, the LOS probability model is formulated as a close-formed expression that can be used at any height of the transceiver. \emph{Fourth}, we develop a resultant path loss model for A2A channels and conduct numerical simulations, 
which provide a clear understanding of 
the relationship between path loss and link distance. In particular, we incorporate the wobbling effects that lead to an imperfect antenna pattern alignment into the analysis of path loss.

The remainder of this paper is organized as follows. Section II first presents the theoretical background of Fresnel zone-assisted channel modeling. Section III systematically describes the methodology of A2A channel modeling, in terms of LOS probability, and path loss model. Numerous results are included in Section IV. We conclude the paper in Section V.


\section{Preliminary Analysis}
In this section, we will first introduce the Fresnel zone of wave propagation, which will facilitate modeling the large-scale channel fading. Then, we introduce the general urban model with two parameters to describe the built-up area. 

\subsection{Fresnel Zones of Electromagnetic Wave Propagation}

The impact of an obstacle can be assessed qualitatively and intuitively by the concept of Fresnel zones that are composed of a series of ellipsoids, as shown in Fig.~\ref{fresnel}. More specifically, the $n$th Fresnel ellipsoid is the one that results in a phase shift of $n\pi$, as compared to the LOS path. The radius of the $n$th Fresnel zone is expressed as
\begin{equation}\label{Radius}
r_n=\sqrt{\frac{n\lambda d_1 d_2}{d_1+d_2}},
\end{equation}
where $d_1$ and $d_2$ represent the distance between the obstacle and transmitter (Tx) and receiver (Rx), respectively. It shows that the maximum radius is when $d_1=d_2=d/2$ where $d$ is the horizontal distance between Tx and Rx. Thus, the minor axis $r_n^{\text{max}}$ of the ellipsoid can be calculated as
\begin{equation}
r_n^{\text{max}}=\frac{\sqrt{n\lambda d_{\text{LOS}}}}{2}.
\end{equation}
Moreover, the focal length of the ellipsoid is $\frac{d_{\text{LOS}}}{2}$ and $d_{\text{LOS}}=\sqrt{d^2+(h_t-h_r)^2}$ where $h_t$ and $h_r$ are the heights of Tx and Rx, respectively. Herein $r_1^{\text{max}}$ is small due to tiny wavelengths at millimeter-wave (mmWave) bands. For $f=28$~GHz, $d_{\text{LOS}}\in(0, 300)$~m, the maximum radius $r_1^{\rm{max}} \in (0, 0.89)$~m.

\subsection{Description of Built-Up Environment}
For mathematically describing the built-up areas, we introduce a simple model that takes the density and height of the buildings into consideration. Herein, we use $\beta_h$ to represent the number of buildings per square kilometers, and $\sigma_h$ is the parameter of the Rayleigh distributed building height. 
The building height 
in the selected environment follows the Rayleigh distribution according to the proposal of ITU-R \cite{x12}, where its probability density function (PDF) is expressed as
\begin{equation}
f(h_b)=\frac{h_b}{\sigma_h^2}\exp(-\frac{h_b^2}{2\sigma_h^2}),
\end{equation}
where $h_b$ is the height of building, and the average height of buildings is $E(h_b)$=$\frac{\sqrt{2\pi}}{2}\sigma_h$. 

\subsection{Preliminary Analysis}
As shown in Fig.~1, the heights of the transceivers are $h_t$ and $h_r$. There are $n$ Fresnel zones with the same two foci and the major axis. For the LOS probability modeling, since the radius of the first Fresnel ellipsoid is much lower than the heights of buildings and transceivers, we calculate the LOS probability as the probability that the height of a building exceeds the direct path. For large-scale channel modeling, the path loss model is based on the two-ray model in the LOS condition, and the path loss mainly contains the Fresnel-Kirchhoff diffraction loss and the free-space propagation loss in the NLOS condition \cite{x13}. The resultant path loss can be obtained according to the LOS probability and the corresponding path loss model in different propagation conditions. 

\begin{figure}[!t]
  \centering
  \includegraphics[width=3.2in]{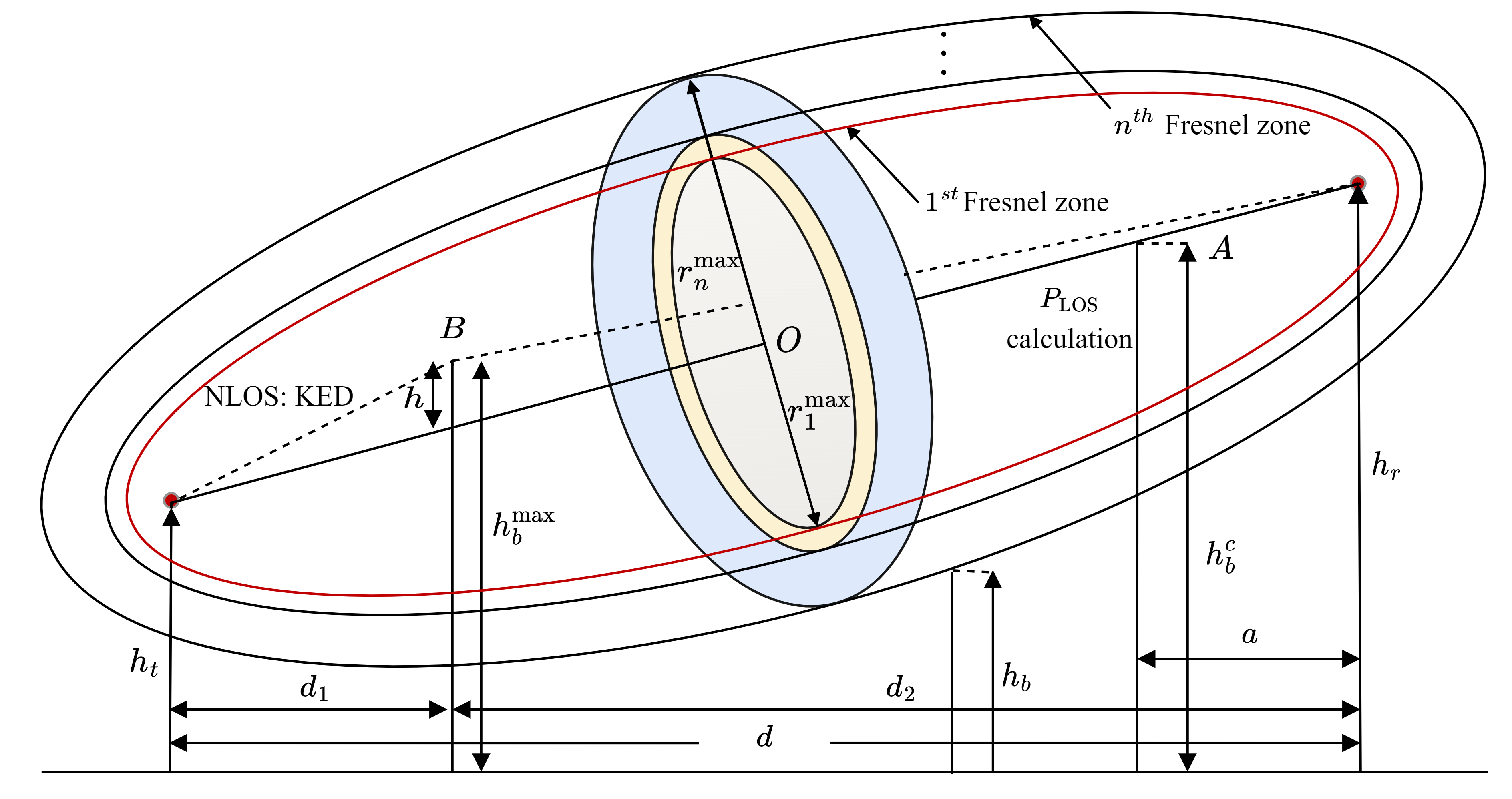}
  \caption{Illustration of A2A wave propagation Fresnel zones in built-up areas.}
  \label{fresnel}
 \end{figure}
\section{A2A Path Loss Modeling }
\subsection{LOS Probability Modeling}
It is critical to determine the LOS probability of an A2A link in an urban environment, even though the probability is generally high at high altitudes. However, for dense urban scenarios and flights in low altitudes, the direct path is readily blocked by buildings such as skyscrapers. To this end, we introduce the wave propagation Fresnel zone, which is also applied to the actual deployment of a terrestrial base station (BS), where a certain clearance of the first Fresnel zone should be guaranteed for conductive signal transmission \cite{x14}.

Figure~\ref{fresnel} illustrates the wave propagation of A2A links based on Fresnel zones, among which the first zone is used for determining the LOS probability. However, it can be found that the first Fresnel zone shaped as an ellipsoid is quite narrow at mmWave bands, which facilitates us to use a more tractable way to determine the LOS probability. 
Therefore, we simplify the calculation of LOS probability, focusing on the line between the Tx and Rx without considering the minor axis of the Fresnel ellipsoid. In other words, the LOS probability can be obtained by determining the probability that the height of a building is lower than the line and then incorporating the number of buildings. Mathematically, we denote the critical height when the building just \emph{touches} the line as $h_b^{c}$ (as shown in point $A$ of Fig.~\ref{fresnel}), which can be determined by
\begin{equation}
    \begin{split}
   h_b^{c}&= |h_t-h_r|+(d-a)\tan(\theta) \\
                 &= \frac{(d-a)h_r+ah_t}{d}=s h_t + (s-1) h_r,
     \end{split}
\end{equation}
where $\theta$ is the elevation angle between Tx and Rx, and $a$ represents the distance between the building and Tx or Rx, and is uniformly distributed in $(0, d)$ and thus $s=\frac{a}{d}$ following $\mathcal{U}(0, 1)$. Finally, the LOS probability considering one typical building can be expressed in the following lemma.

\textbf{Lemma 1:} \emph{Consider one typical building, the LOS probability can be calculated in terms of the Rayleigh parameter and the heights of transmitter and receiver, expressed as}
\begin{equation}
P_\text{LOS}^\text{T}=
\begin{cases}
&1-\exp\left(-\frac{H^2}{2\sigma_h^2}\right),\;\text{for}\;h_t = h_r,\\
&1-\sqrt{\frac{\pi}{2}}\sigma_hH_d,\;\;\;\;\;\;\text{for}\;h_t \neq h_r,
  \end{cases}
  \label{eq.plos}
\end{equation}
\emph{where $H=h_t= h_r$ and  $\text{erf}(x)=\frac{2}{\sqrt{\pi}}\int_{0}^{x}\exp(t^2)dt$ represents the Gaussian error function and $H_d$ is expressed as}
\begin{equation}
H_d=\frac{\text{erf}\left({\frac{h_t}{\sqrt{2}\sigma_h}}\right)-\text{erf}\left({\frac{h_r}{\sqrt{2}\sigma_h}}\right)}{h_t-h_r}.
\end{equation}
\emph{Besides, the LOS probability is a continuous function in terms of the independent variables since the limitation of the break-point in the condition of $h_\text{t} \neq h_\text{r}$ is equal to the result of the condition of $h_\text{t} = h_\text{r}$.}

\emph{Proof:} Since the height distribution of buildings follows Rayleigh distribution, the LOS probability considering one building can be calculated by 
\begin{equation}
	\begin{split}
P_\text{LOS}^\text{T}&= \text{Pr}(h_b < h_b^{c})\\
	&= \int_{0}^{1}\int_{0}^{s h_t + (s-1) h_r}f(h_b)dhds\\
    &=1-\sqrt{\frac{\pi}{2}}\sigma_h \frac{\text{erf}\left({\frac{h_t}{\sqrt{2}\sigma_h}}\right)-\text{erf}\left({\frac{h_r}{\sqrt{2}\sigma_h}}\right)}{h_t-h_r}.
	\end{split}
\end{equation}
where $h_\text{t} \neq h_\text{r}$. For the condition of $h_r= h_t$, the critical height $h_b^{c}=H$, thereby the LOS probability can be determined by
\begin{equation}
	\begin{split}
P_\text{LOS}^\text{T}&=\int_{0}^{H}f(h_b)dh=1-\exp\left(-\frac{H^2}{2\sigma_h^2}\right).
	\end{split}
\end{equation}
For continuity, we have to prove the limitation of the break-point in (6) is identical to (8). Without loss of generality, we denote $h_t$ as an independent variable and $h_r$ as a constant. Thus, the limitation for $h_t \to h_r$ can be calculated by
 \begin{equation}
 \begin{split}
&\lim_{h_t \to h_r} H_d \overset{\Delta}{=}\lim_{h_t \to h_r}\frac{\frac{d}{dh_t}\left(\text{erf}\left({\frac{h_t}{\sqrt{2}\sigma_h}}\right)-\text{erf}\left({\frac{h_r}{\sqrt{2}\sigma_h}}\right)\right)}{\frac{d}{dh_t}(h_t-h_r)}\\
&=\lim_{h_t \to h_r} \frac{\sqrt{\frac{2}{\pi}}\exp\left(-\frac{h_t^2}{2\sigma_h^2}\right)}{\sigma_h}=\frac{\sqrt{\frac{2}{\pi}}\exp\left(-\frac{H^2}{2\sigma_h^2}\right)}{\sigma_h}.
	\end{split}
\end{equation}
where ($\Delta$) applies the l'H\^{o}pital's rule. By combining (7) and (9), we can obtain (8), and thus the continuity of the LOS probability is proved. $\hfill\blacksquare$

\begin{figure}[!t]
  \centering
  \includegraphics[width=3in]{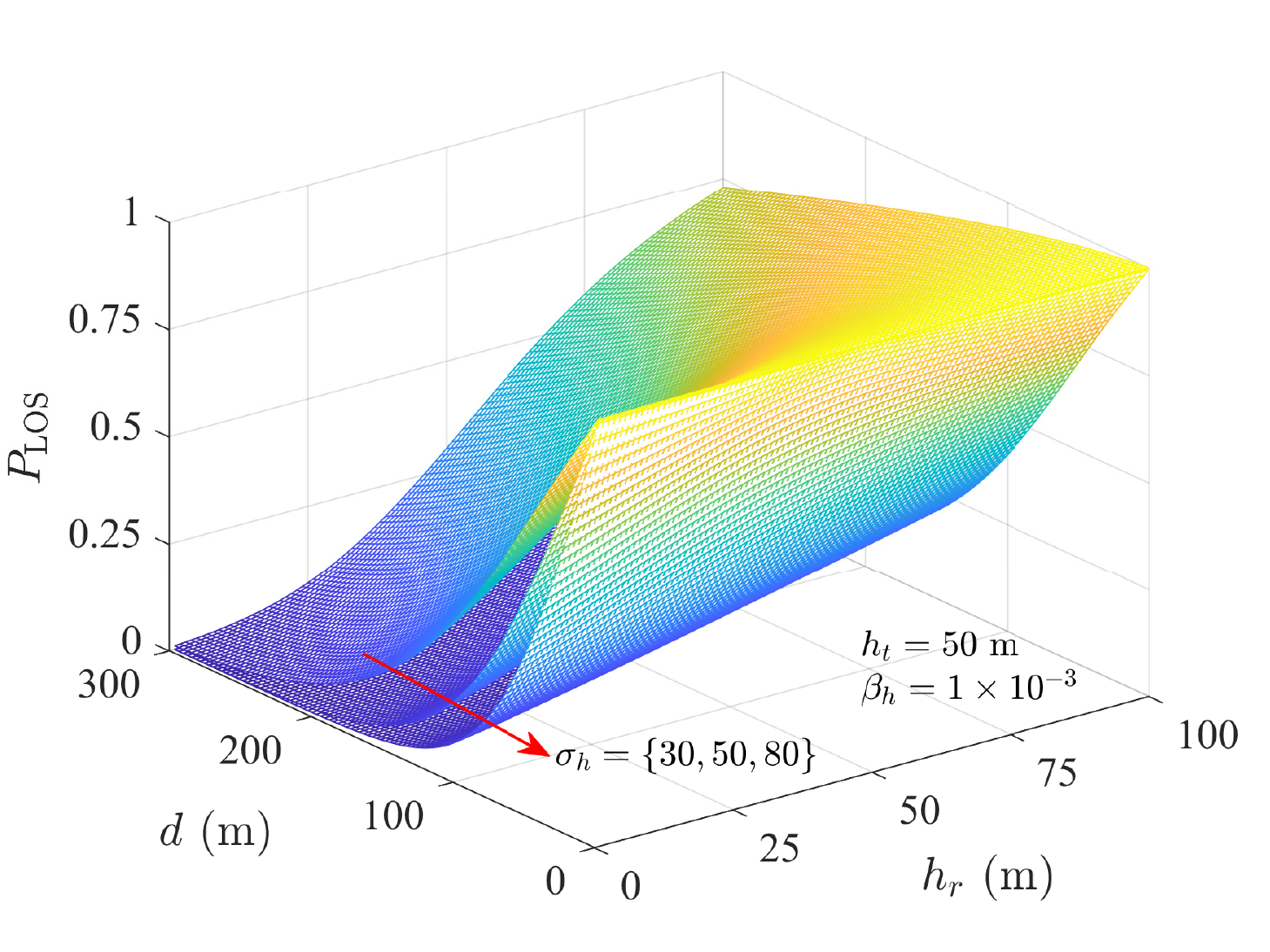}
  \caption{Proposed LOS probability as a function of $d$, $h_r$, and $\sigma_h$.}
  \label{fig.Plos}
 \end{figure}
 
It intuitively shows that the LOS probability is a function of the 
building heights as well as the 
transceiver altitudes. As an example, in an open environment that is free of buildings, i.e., $\sigma_h=0$, the LOS probability is always 1 regardless of the height of the transceiver. More specifically, the LOS probability degrades as $\sigma_h$ increases.

To determine the LOS probability considering the buildings that may cross the line, we need to define the expected number of buildings. Considering $\beta_h$ is an area density, we denote the area of the ground projection of the first Fresnel zone as $S$ where $S\approx\frac{\pi d}{2}r_1^{\max}$. Thus, the average number of buildings is $E(b)=S\beta_h$. Then, the closed-form LOS probability can be expressed as
\begin{equation}
 P_\text{LOS}=(P_\text{LOS}^\text{T})^{E(b)}=(P_\text{LOS}^\text{T})^{S\beta_h}.
\end{equation}
As shown in Fig.~\ref{fig.Plos}, we can verify that the LOS probability is relatively large when the horizontal distance ($d$) is short, which is in accordance with the results in standard technical reports, such as 3GPP \cite{x15} and WINNER II \cite{x16}. As an example, the 3GPP report suggests that the LOS probability is 1 when the distance is smaller than 18~m for an urban environment. Moreover, the LOS probability remains 1 when the height of the aerial user is higher than 100~m where the height of the ground cellular base station is 25~m or 10~m for macro or micro types. According to Eq.~(\ref{eq.plos}), we can verify that $H_d$ will be very small for the large difference of height between Tx and Rx since $|\text{erf}(h_t/\sqrt{2}\sigma_h)-\text{erf}(h_r/\sqrt{2}\sigma_h)|\le 1$ holds.

\subsection{Imperfect Beam Alignment}


In mmWave bands, it is important to incorporate the directional antenna model since the instantaneous directive gain can remedy the large path loss. The 
resulting antenna gain $G=G_t(\theta_{\text{tx}})G_r(\theta_{\text{rx}})$,where $G_t(\theta_{\text{tx}})$ given by \cite{x17},
\begin{equation}
G_t(\theta_{\text{tx}})=N\cos\left(\frac{\pi N}{2}(\theta_{\text{tx}}+\theta_{\text{tx}}^{w})\right)^m, \theta_{\text{tx}}\le\frac{1}{N},
\end{equation}
where $N$ represents the number of antennas in the uniform linear array (ULA) for the Tx. The value of $m$ is set to 2 for analytical tractability in \cite{x17}. Due to the instability of the UAV flight, the corresponding fading behavior is also highly influenced, as reported in \cite{x18}. Therefore, it is necessary to consider the wobbling effects in the A2A channel modeling. Thus, we denote the fluctuating angles as $\theta_{\text{tx}}^{w}$ and $\theta_{\text{rx}}^{w}$ for Tx and Rx, respectively. Generally, these angles are modeled as Gaussian random variables (RVs) with zero mean and small variations, expressed as $\theta_{\text{tx, rx}}^{w} \sim \mathcal{N}(0,\sigma_{\text{tx, rx}}^2)$. In this paper, we consider the beams of Tx and Rx to be well-aligned based on tracking algorithms, thus, there is $\theta_{\text{tx}}=\theta_{\text{rx}}=0$.
\subsection{Path Loss Modeling}
Path loss and shadow fading consist of large-scale fading that occurs at a long distance in terms of wavelength, generally in several hundred wavelengths. We herein investigate the path loss in both LOS and NLOS conditions.
\subsubsection{LOS Condition}
The popular path loss model used in A2A channels is the free-space (FS) model derived from the Friis' law, which can be expressed as $PL_{\text{Friis}}=G\left(\frac{\lambda}{4\pi d_{\text{LOS}}}\right)^2$. Since it is the LOS condition of the A2A link, we consider a ground reflection, which consists of a more intuitive two-ray model, given by
\begin{equation}
PL_{\text{LOS}}=G\left(\frac{\lambda}{4\pi d_{\text{LOS}}}\right)^2|1-P_{\rm GR}\Gamma\exp(j\Delta \phi)|^2,
\end{equation}
where $G=G_tG_r$ represents the total antenna gain, $\Delta \phi=\frac{2\pi}{\lambda}(\sqrt{(h_t+h_r)^2+d^2}-\sqrt{(h_t-h_r)^2+d^2})$ represents the phase difference between the LOS and ground reflection, and $\Gamma$ is the reflection coefficient assumed to be 1. $P_{\rm GR}$ is the probability of ground reflection, referring to the probabilistic two-ray model in \cite{x2}. Moreover, $P_{\rm GR}$ can be obtained by the proposed LOS probability model, which is expressed as 
\begin{equation}
P_{\rm GR}=P_{\rm LOS}(d_t, h_t, 0)\cdot P_{\rm LOS}(d_r, h_r, 0),
\end{equation}
where $d_t$ and $d_r$ are the horizontal distances between the ground reflected point and Tx and Rx, respectively, which can be obtained by $d_{t}=\frac{dh_{t}}{h_t+h_r}$ and $d_{r}=\frac{dh_{r}}{h_t+h_r}$. Note that due to the small size of the first Fresnel zone with $d_{\text{LOS}} \gg \lambda$, there is no multipath in the first Fresnel zone, which results in the path loss as a function of $d_{\text{LOS}}^{-2}$.

\subsubsection{NLOS Condition}
For a NLOS condition, at least one building blocks the direct path between Tx and Rx. However, we only need to consider the diffraction produced by the tallest building. Thus the propagation becomes a typical knife-edge diffraction (KED), as shown in Fig.~\ref{fresnel} (point $B$). Refer to \cite{x19}, the diffraction loss is given by
\begin{equation}
L(v) [\text{dB}]=6.9+20\log10(\sqrt{(v-0.1)^2+1}+v-0.1),
\end{equation}
where 
$v\ge-0.78$ is the Fresnel-Kirchhoff diffraction parameter, given by 
\begin{equation}
v=h\sqrt{\frac{2(d_1+d_2)}{\lambda d_1 d_2}}\ge h\sqrt{\frac{8}{\lambda d}},
\end{equation}
where $h$  is the height of the top of the obstacle above the straight line joining the two ends of the path. 
For the analytical simplicity, we herein consider the lower bound of $v$, i.e, $v_{\min}=h\sqrt{\frac{8}{\lambda d}}$, with $d_1d_2\le\left(\frac{d_1+d_2}{2}\right)^2=\frac{d^2}{4}$.

The diffraction loss constitutes the dominant loss in the total path loss in the NLOS condition. Moreover, the diffraction from the highest building is the most important, since multipath components (MPCs) from other buildings will suffer from the penetration loss from the highest building, these MPCs are negligible in mmWave bands. For characterizing the diffraction from the highest buildings, we need first to derive the distribution of the highest height of building that is denoted as $h_b^{max}$. Because the heights of buildings are independent and identically distributed (i.i.d.), the cumulative distribution function (CDF) of $h_b^{max}$ can be expressed as
\begin{equation}
\begin{aligned}
F_{h_b^{max}}(x)&=Pr(h_b^{max}\le x)\\&=Pr(h_b^{1}\le x)\cdot Pr(h_b^{2}\le x)\cdot \cdot \cdot Pr(h_b^{N}\le x)\\&=\left(1-\exp\left(-\frac{x^2}{2\sigma_h^2}\right)\right)^N,
\end{aligned}
\end{equation}
where $N$ is the number of buildings within the projection. 

Without the loss of generality, we assume $h_r\ge h_t$, the expectation of $h$ in the NLOS condition can be calculated by
\begin{equation}
\begin{aligned}
E(h) &=E\left(h_b^{max}-\frac{d_1}{d}(h_r-h_t)-h_t\right),\\
&=E\left(h_b^{max}-\frac{d_1}{d}h_r-\frac{d-d_1}{d}h_t\right),\\
&=E(h_b^{max})-\frac{h_r}{2}-\frac{h_t}{2},
\end{aligned}
\end{equation}
where $d_1$ follows the uniform distribution  in (0, $d$) and thus $E(d_1)=d/2$. As an example, for $h_t=h_r=H$, $E(h)=E(h_b^{\max})-H$. In addition, $E(h_b^{max})$ can be calculated by
\begin{equation}
\begin{aligned}
E(h_b^{\max}) &=\int_{\min\{h_t,h_r\}}^{\infty} 1- F_{h_b^{\max}}(x) dx,\\
&=\sum_{n=1}^{N}(-1)^{n-1}\frac{C_{N}^{n}}{n}\sigma_h \sqrt{n\pi}\;\text{erfc} \left(\frac{h_t\sqrt{n}}{2\sigma_h}\right),
\end{aligned}
\end{equation}
where the combination is denoted as $C_{N}^{n}=\frac{N!}{n!(N-n)!}$. As an example, when $N=2$, $E(h_b^{\max}) $ can be expressed as
\begin{equation}
E(h_b^{\max}) =\frac{\sigma_h \sqrt{\pi}}{2}\left(4 \text{erfc}\left(\frac{h_t}{2 \sigma_h}\right) - \sqrt{2} \text{erfc}\left(\frac{h_r}{\sqrt{2} \sigma_h} \right)\right).
\end{equation}
It is confirmed that the two-ray model is not effective when a building blocks both the direct and reflected paths. Thus, the path loss in NLOS condition can be obtained by the addition of the free-space loss and diffraction loss, given by
\begin{equation}
PL_{\text{NLOS}} ~[\text{dB}]=10\log(PL_{\text{Friis}})+L(v).
\end{equation}

Therefore, the total path loss considering the link state probability can be expressed as 
\begin{equation}
PL ~[\text{dB}]=P_{\text{LOS}}PL_{\text{LOS}}+(1-P_{\text{LOS}})PL_{\text{NLOS}}.
\end{equation}







\begin{figure}[!t]
  \centering
  \includegraphics[width=3in]{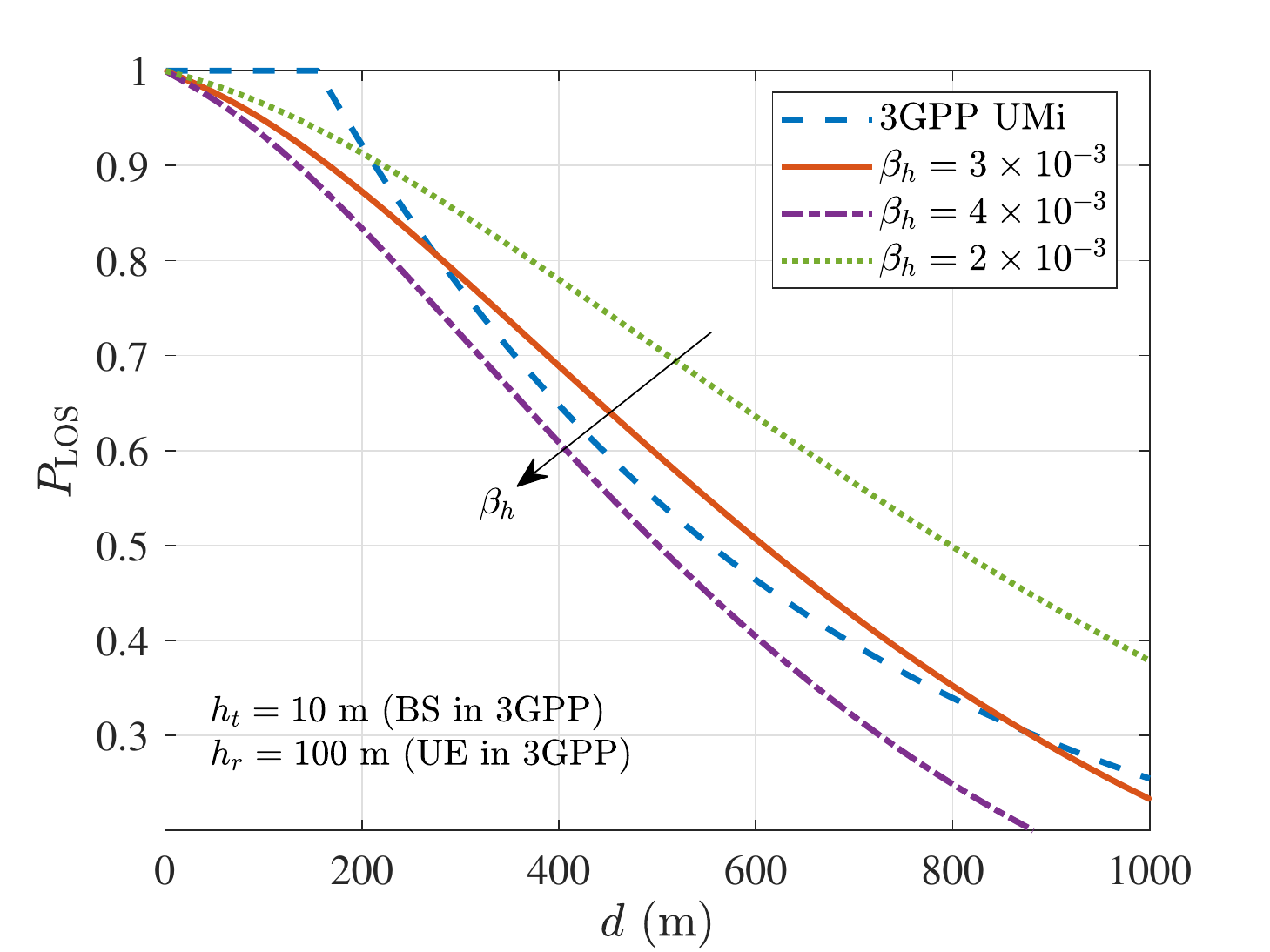}
  \caption{Comparison of LOS probability in urban environment when $\sigma_h=20$.}
  \label{Plos-vali}
 \end{figure}
 \begin{figure}[!t]
  \centering
  \includegraphics[width=3in]{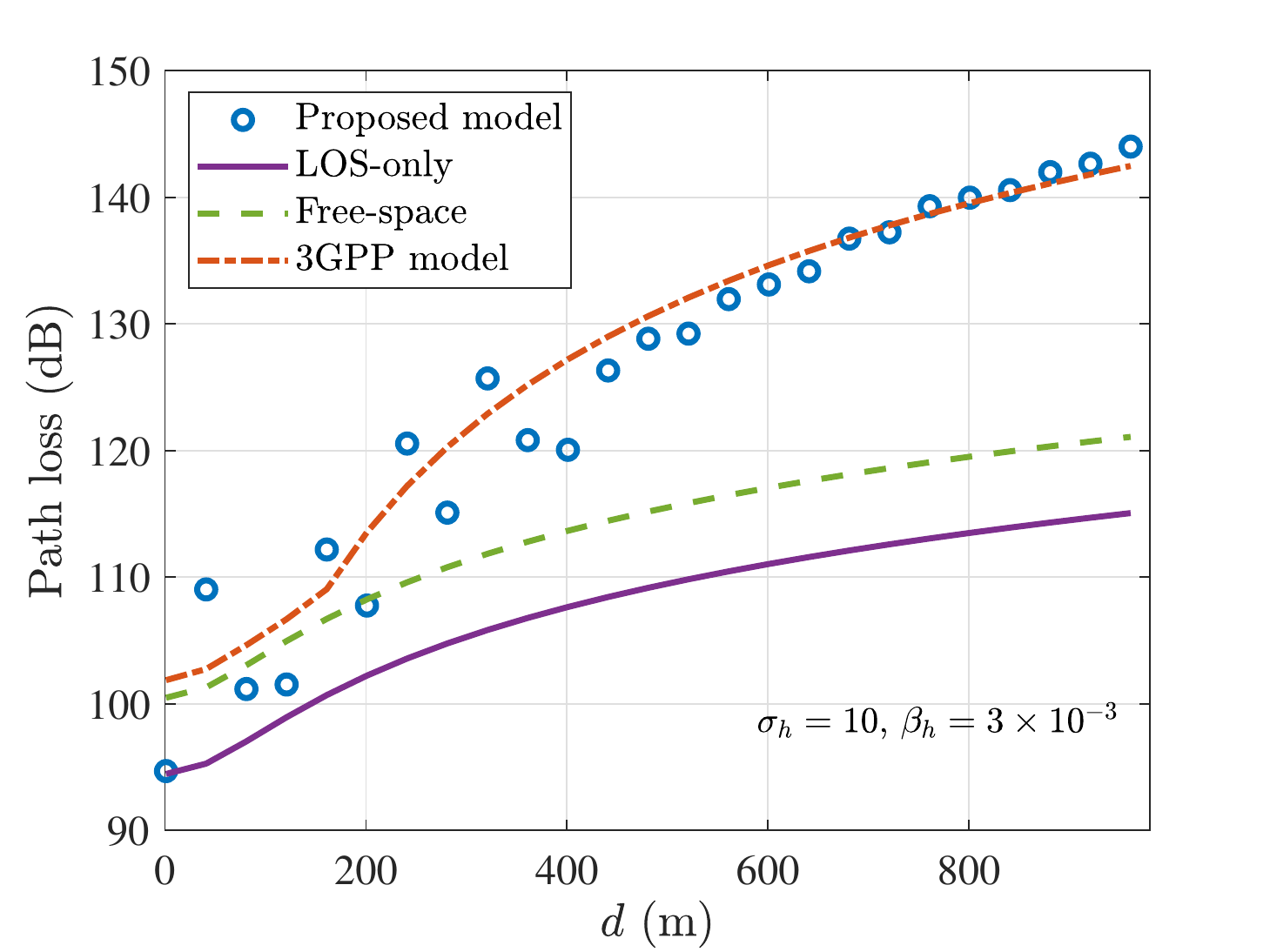}
  \caption{Path loss comparisons with 3GPP, LOS-only, and free-space models.}
  \label{PL-vali}
 \end{figure}
\section{Results }
We will first compare the proposed path loss model with the 3GPP model to validate the model. Then, we will show the path loss results in terms of different urban environments as well as the distance/height of the UAVs. Finally, the impact of imperfect beam alignment on the path loss fluctuations (termed shadow fading) will be investigated, which will provide an empirical relationship between the beam misalignment level and path loss fluctuations. In the analysis below, the carrier frequency is set at 28~GHz. It is noted that other simulation parameters may vary based on different analytical purposes. 
\subsection{Comparison of Proposed Model}
In the 3GPP 36.777 model \cite{x15}, the LOS probability in an urban micro (UMi) scenario is a function of the height of UE and the distance between UE and BS. More specifically, in the 3GPP, $P_{\rm LOS}=1$ when $d\le d_0$, and $P_{\rm LOS}=\frac{d_0}{d}+\exp\left(-\frac{d}{p_1}\right)\left(1-\frac{d_0}{d}\right)$ when $d> d_0$, where $d_0=\max(18,294.05\log10(h_r)-432.94)$ and $p_1=233.98\log10(h_r)-0.95$. Note that the height limit of UE in the 3GPP is 300~m, which corresponds to our scope of low altitude. We compare the proposed LOS probability model with the 3GPP model as a function of $d$, as illustrated in Fig.~\ref{Plos-vali}. We can find that the 3GPP model is independent of environmental parameters, such as $\sigma_h$ and $\beta_h$ we used in our model. The result shows that $P_{\rm LOS}$ results of the proposed and 3GPP models present a similar trend when $\beta_h=3\times10^{-3}$ and $\sigma_h=20$. However, the 3GPP model is incapable of extending to other urban environments with different densities and heights of buildings, which suggests our proposed model tackles the limitation of environmental specificity in channel modeling, thus resulting in better extensibility.

Then, we employ the LOS probability in determining the total path loss and compare the results between the proposed model and the 3GPP path loss model, where we consider $h_t=10$~m, $h_r=100$~m, $\sigma_h=10$, and $\beta_h=3\times10^{-3}$. 
Fig.~\ref{PL-vali} illustrates that the free-space and LOS-only models have large differences from the proposed and 3GPP models, which infers that the actual wave propagation is more complex than LOS-only/free-space propagation. Nonetheless, both the proposed model and 3GPP model show that the propagation is close to the free space when $d$ is short, e.g., $d\le150$~m and $d\le200$~m for 3GPP and the proposed model, respectively, shown in Fig.~\ref{PL-vali}. 
When $d$ increases, the results between the proposed model and the 3GPP model agree well. However, our model may present a step-wised trend due to the integral of Eq.~(18) which is as a function of $N$ (resulting in a discrete $v$). When $d \le 200$~m, the path loss is indeed well approximated by 
the free-space loss, then at $d > 200$ m the loss increases due to the low LOS probability and high diffraction loss, as predicted also by our proposed model.

\subsection{Path Loss Results}
\begin{figure}[!t]
  \centering
  \includegraphics[width=3in]{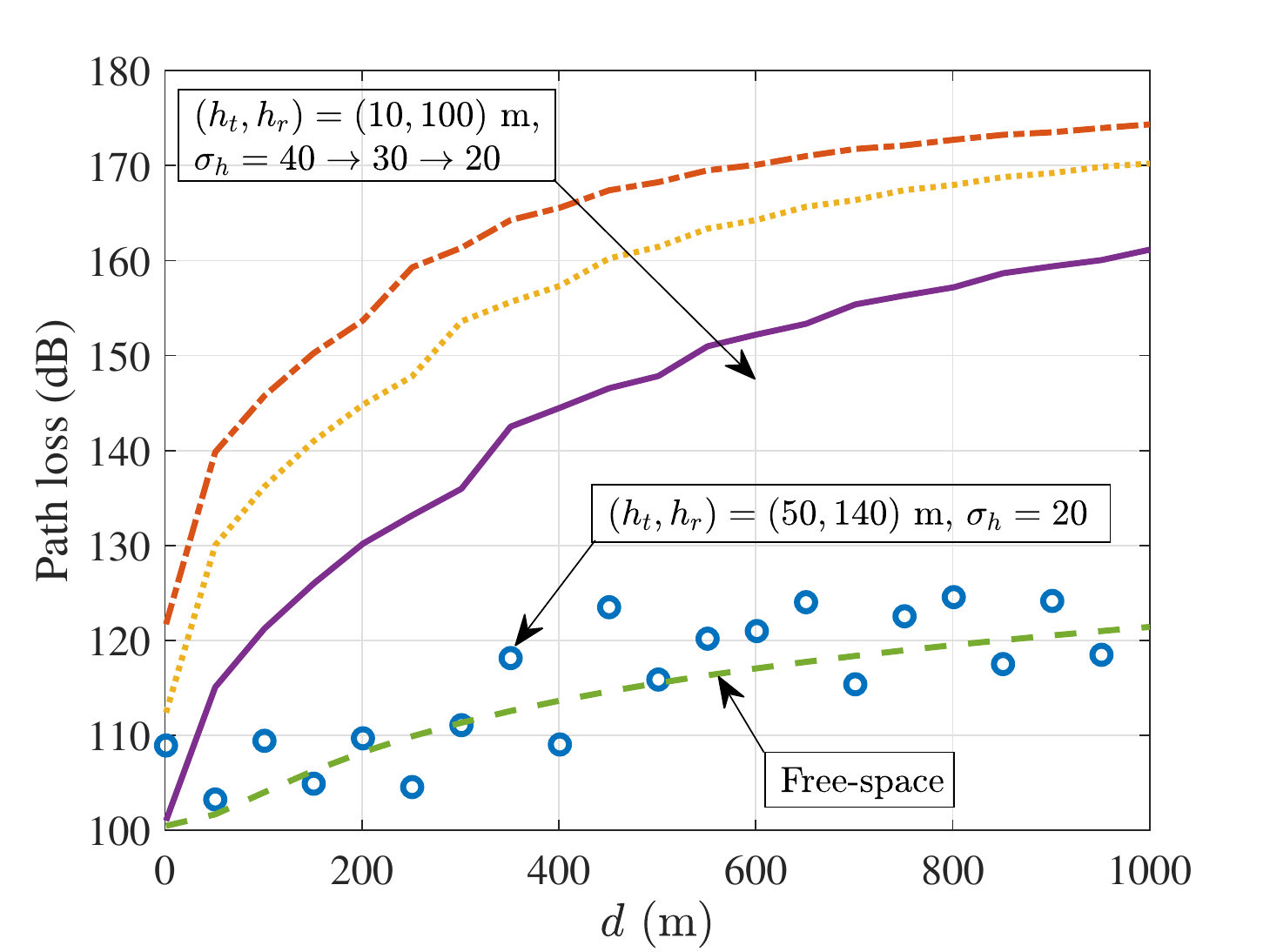}
  \caption{Path loss results for different ($h_t$, $h_r$, $\sigma_h$) as a function of $d$.}
  \label{Pl-result}
 \end{figure}

We then investigate the path loss for different height pairs of Tx and Rx ($h_t$, $h_r$), and environmental parameters ($\beta_h$, $\sigma_h$) where $\sigma_h=20,30,40$ with corresponding $\beta_h=\{2,3,4\}\times10^{-3}$ denote suburban, urban, and dense urban, respectively. As 
the denser urban environments lead 
to the higher path loss, shown in Fig.~\ref{Pl-result}, which can be explained by an increasing $N$ resulting in a higher $E(h_b^{\max})$ for an increased density. 

We then change the 
Tx and Rx heights at $\sigma_h=20$, while keeping the 3D distance $d_{\rm LOS}$ constant. Fig.~\ref{Pl-result} shows that the higher UAV altitudes 
 result in 
a lower path loss because of the higher LOS probability. As an example, we can find that, for $h_t=50$~m, the path loss agrees well with the free-space path loss for the considered horizontal distance (max. 1000~m). In this environment, an empirical finding is when both UAVs are higher than 50~m, the path loss of A2A channels can be described with a free-space propagation model.
 
\subsection{Impact of Imperfect Beam Alignment}
\begin{figure}[!t]
  \centering
  \includegraphics[width=3in]{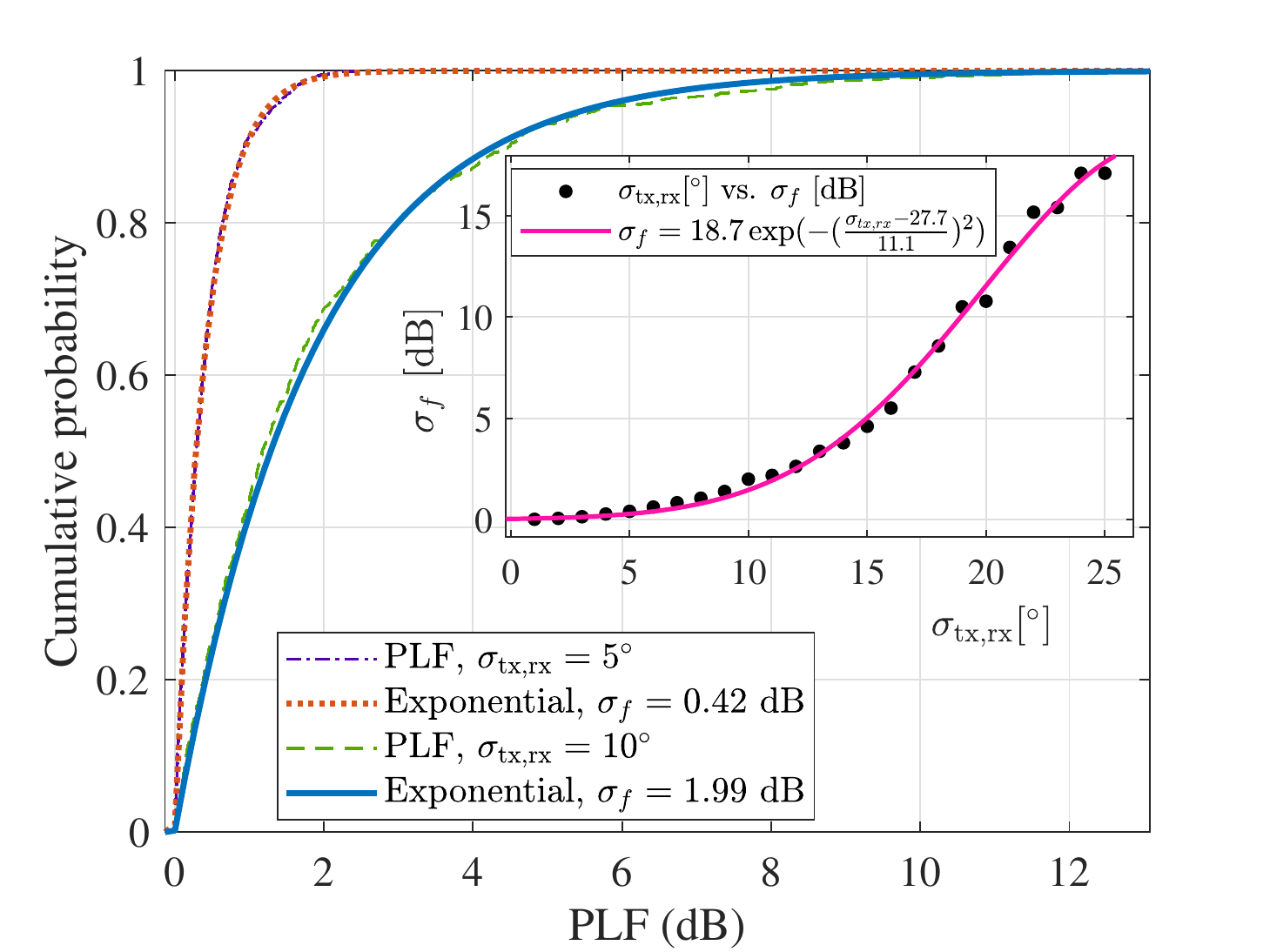}
  \caption{CDFs of path loss fluctuations, and illustration of the relationship between PLF and beam alignment level, i.e., $\sigma_f$ versus $\sigma_{\rm tx,rx}$.}
  \label{shadow-result}
 \end{figure}

To show the impact of beam misalignment level on the path loss model, we first obtain the path loss considering the random antenna gains $G$ as a function of $\sigma_{\rm tx,rx}$. Note that we assume $\sigma_{\rm tx}=\sigma_{\rm rx}$ in the analysis. The fluctuations can be calculated by the difference between the path loss considering a random $G$ and a constant $G=1$, denoted as $\rm PLF$ (dB). As shown in Fig.~\ref{shadow-result}, we can find that the CDFs of $\rm PLF$ follow an exponential distribution. 
It can be theoretically proved that for a Gaussian random variable $\theta_{\rm tx,rx}^f$, its cosine variant follows an exponential-form distribution \cite{x20}. 
The standard deviation of $\rm PLF$ is denoted as $\sigma_f$ (dB), and results show that a more serious misalignment level $\sigma_{\rm tx,rx}$ leads to a higher $\sigma_f$. 

To obtain the relationship between $\sigma_{\rm tx,rx}$ and $\sigma_f$, as shown in an embedded figure of Fig.~\ref{shadow-result}, we found that the exponential fit agrees well with the relation, expressed as $\sigma_f=18.7\exp\left(-\left(\frac{\sigma_{\rm tx,rx}-27.7}{11.1}\right)^2\right).$ Thus, the fluctuations can be quantified with known misalignment levels, which is helpful for planning the link budget of A2A communications.

\section{Conclusion and Future Work}
In this paper, we proposed a path loss model for A2A channels considering a generic built-up environment and different link states. First of all, a closed-form LOS probability was derived, which can be used for 
a range of urban environments and any 
Tx and Rx height. Then, considering LOS and NLOS conditions, a resultant path loss model was developed. A comparison with the 3GPP model shows the effectiveness and high scalability of the proposed model. Numerical analyses illustrate that the A2A channels experience free-space propagation for altitudes higher than 50~m for the considered environment. For ultra-low altitudes, the path loss is highly related to the density of buildings. Finally, we quantified the path loss fluctuations caused by imperfect beam alignment. 
To further extend the analysis, future work can be done in the following aspects: 1) small-scale fading considering $n$ Fresnel zones, and 2) more validations by measurement campaigns.

\section*{Acknowledgement}
This work is partly supported by the Research Foundation Flanders (FWO), project no. G098020N, cooperating with Prof. Wout Joseph at Ghent University, and partly by KU Leuven Postdoctoral Mandate (PDM) under project no. 3E220691.





\end{document}